\documentclass[sigconf]{acmart}

\usepackage[utf8]{inputenc}
\usepackage{subcaption}
\usepackage{balance}
\usepackage{footnote}
\usepackage{url}
\usepackage{float}
\usepackage{tikz}
\usetikzlibrary{positioning}

\definecolor{amethyst}{rgb}{0.6, 0.4, 0.8}
\definecolor{aqua}{rgb}{0.0, 1.0, 1.0}

\renewcommand\footnotetextcopyrightpermission[1]{} 
\title[Data4UrbanMobility]{Data4UrbanMobility: Towards Holistic Data Analytics for Mobility Applications in Urban Regions}

\begin{document}

\author[Tempelmeier et al.]{Nicolas~Tempelmeier, Yannick~Rietz, Iryna~Lishchuk, Tina~Kruegel, Olaf~Mumm, Vanessa~ Miriam~Carlow, Stefan~Dietze, Elena~Demidova}

\affiliation{
 \institution{L3S Research Center, Leibniz Universit\"at Hannover}
}

\affiliation{
\institution{Institute for Sustainable Urbanism, TU Braunschweig}
}

\affiliation{
  \institution{PROJEKTIONISTEN GmbH}
}

\affiliation{
\institution{Institute for Legal Informatics, Leibniz Universit\"at Hannover}
}

\affiliation{
\institution{GESIS - Leibniz Institute for the Social Sciences}
}

\email{
tempelmeier@L3S.de, 
rietz@projektionisten.de, 
iryna.lishchuk@iri.uni-hannover.de, 
kruegel@iri.uni-hannover.de, 
}
\email{o.mumm@tu-braunschweig.de, 
v.carlow@tu-braunschweig.de,
stefan.dietze@gesis.org,
demidova@L3S.de
}

\begin{abstract}
With the increasing availability of mobility-related data, such as GPS-traces, Web queries and climate conditions, there is a growing demand to utilize this data
to better understand and support urban mobility needs. 
However, data available from the individual actors, such as providers of information, navigation and transportation systems, is mostly restricted to isolated mobility modes, whereas holistic data analytics over integrated data sources is not sufficiently supported.
In this paper we present our ongoing research in the 
context of holistic data analytics to support urban mobility
applications in the Data4UrbanMobility (D4UM) project.
First, we discuss challenges in urban mobility analytics and present the D4UM platform we are currently developing to facilitate holistic urban data analytics over integrated heterogeneous data sources along with the available data sources. 
Second, we present the MiC app - a tool we developed to complement available datasets with intermodal mobility data (i.e. data about journeys that involve more than one mode of mobility) using a citizen science approach. 
Finally, we present selected use cases and discuss our future work. 
\end{abstract}

\maketitle

\section{Introduction}
\label{sec:intro}

Contemporary urban mobility behavior is undergoing a rapid transition and paradigm shift and also is affected by a wide range of short-, medium- as well as long-term factors. These vary from aspects such as immediate climate conditions, regional construction sites or ephemeral events to long-term trends such as increasing negative environmental impacts and the widespread adoption of intermodal mobility chains (i.e. journeys involving several means of transportation and mobility such as walking, cycling, public transportation, etc.). 
Moreover, the lifestyle- and population-induced new demand for mobility leads to region-specific, ecological and traffic related problems in growing metropolitan areas and is a limiting factor for urban development.

Traditional urban traffic planning relies on complex models, combining trip generation, trip distribution, mode choice and route choice \cite{McNally:2008} to simulate travel demands. Data foundations of these models essentially consist of traffic census, socio-demographic data as well as infrastructural and service data, where large-scale and accurate data about actual mobility behavior, in particular in the context of intermodality, is costly to obtain and the aforementioned contextual factors are largely ignored. 

Throughout the last decade, the widespread use of Web and mobile applications has led to an increasing availability of data, which captures both actual mobility needs and usage as well as contextual information about, for instance, traffic incidents, city events and weather conditions. This data has the potential to complement existing data sources and information systems currently used for handling urban mobility processes. In particular, within densely populated areas, the correlation of mobility behavior with data obtainable from mobility apps, public transportation websites and social media streams can uncover more complex dependencies and aid the development of supervised models which consider a wide variety of features and enable predictions of future needs.

Despite an overall increasing availability of datasets related to mobility behaviour, 
this data typically focuses on one mode of transportation, most prominently covering individual traffic and, to some extent, public transportation services. Other modes of mobility that become particularly important in modern cities, e.g. cycling and walking, are typically not captured at scale. Furthermore, data sources coming from particular mobility services and transportation providers do not adequately capture intermodal mobility sequences. However, such data is of utmost importance to better understand and predict the actual demand in mobility and associated services.

The challenges related to collection and analysis of such data are manifold. They include  
the provision of tools and methods to capture and analyse intermodal mobility, 
designing incentives for city inhabitants to share their mobility data, protection of 
personal data to be collected in accordance to the legal framework 
as well as data analytics methods and models to provide added value 
for the individual participants, mobility service providers and city authorities. 

In this work, we present our ongoing research within the \linebreak \textit{Data4UrbanMobility} project (D4UM)\footnote{\url{http://data4urbanmobility.l3s.uni-hannover.de/}}. 
This research aims at gathering, augmenting and analysing mobility data in urban regions to address the problems presented above.
We introduce the D4UM platform built to facilitate collection and analysis of mobility-related data from a variety of sources on a long term.
In the project we in particular focus on the datasets available in the urban regions of Hanover, Wolfsburg and Brunswick (Germany), whereas the results will be transferable to other urban regions. 
The overall contributions of the project include a large annotated data catalogue including regionally and globally relevant mobility-related data sources, models built upon these data sources and analytic results in the particular regions. 

Furthermore, the paper introduces the MiC-App - a human-centric data tool developed in the project framework to capture individual movements – tracks and modes – to complement existing sources in the context of intermodal urban mobility. Combined with other available data sources, the data collected through the MiC-App and the underlying D4UM platform facilitates the capturing and the analyses of data to better understand the growing demand in intermodal mobility in urban regions.

\section{Problem Description}
\label{sec:problem}

The aim of the D4UM project is to facilitate efficient analytics and data-driven estimates of mobility demand and evaluation of mobility network quality, addressing the needs of different stakeholders. The stakeholders of the project include individual citizens, city administrations, mobility service providers and urban traffic planners.
The aims of these stakeholders can be categorised with respect to the different time horizons, as follows: 

\begin{itemize}
    \item Short Term: Facilitate efficient access to mobility services, taking into account temporary high load (citizens).
    \item Medium Term: Provide new and optimise existing mobility services (mobility service providers).
    \item Long Term: Facilitate data-driven integrated urban planning processes (urban traffic planers, city administrations).
\end{itemize}

To address these problems the project develops the D4UM platform that facilitates collection and  
analytics of relevant data sources. 
The research questions that can be addressed using this platform include e.g.:

\begin{itemize}
    \item Which external conditions such as climate or spatial structures influence the typical movement patterns of city inhabitants?
    \item How are the urban movement patterns influenced by the external factors, such as e.g. weather conditions, given specific urban contexts and mobility options?
    \item How can an increased demand in mobility services be determined and addressed?
\end{itemize}

\begin{figure*}
    \centering
    \includegraphics[width=0.6\textwidth]{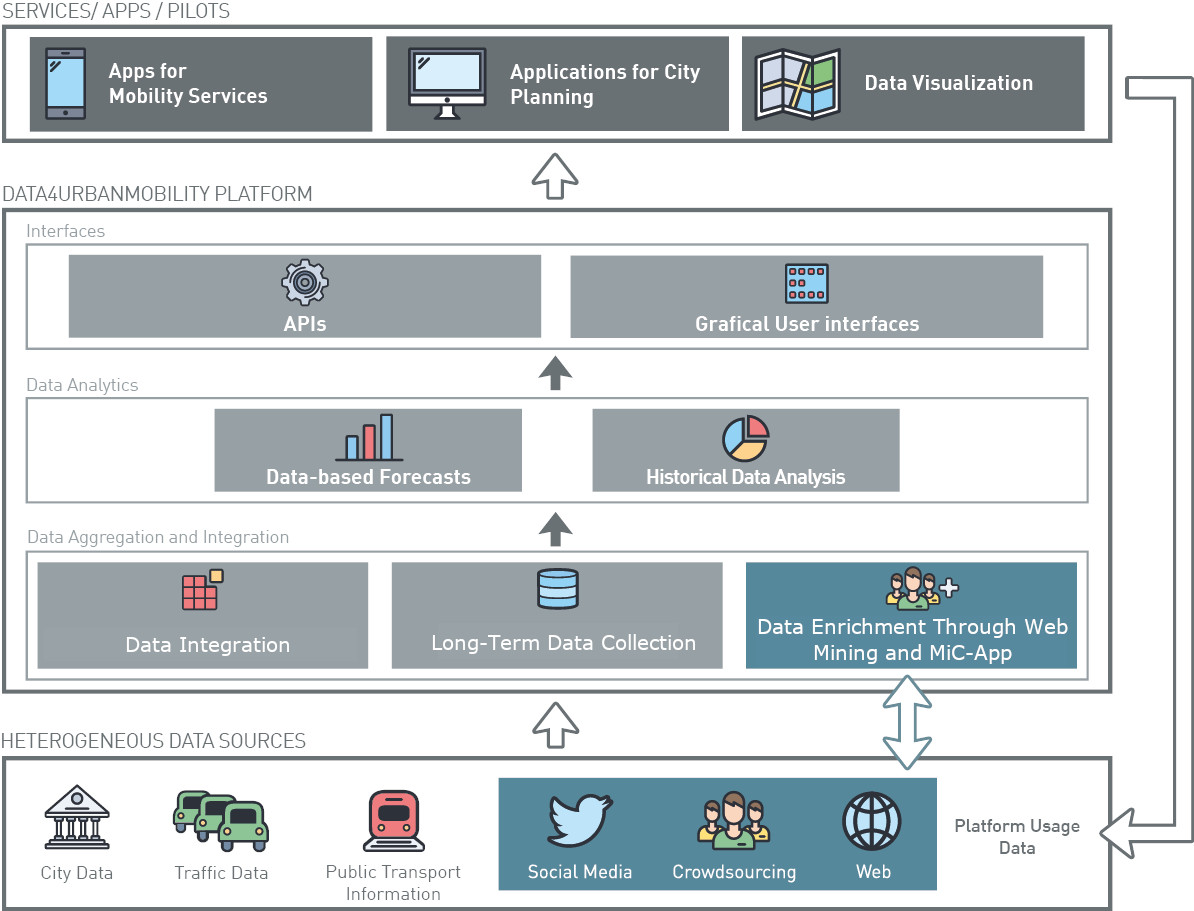}
    \caption{Overview of the Data4UrbanMobility platform architecture}
    \label{fig:frameworkOverview}
\end{figure*}

\section{D4UM Platform}
\label{sec:d4umframework}

This section describes the architecture of the D4UM platform. This platform enables integrated analytics of heterogeneous data sources 
for different stakeholders in the context of urban mobility. Figure \ref{fig:frameworkOverview} provides an overview of the platform layers 
and its individual components. 

The input layer of the D4UM platform consists of \emph{heterogeneous data sources} that cover various aspects of urban mobility. 
These data sources are described in Section \ref{sec:data} in more detail. 

The \emph{data aggregation and integration} layer conducts all necessary pre-processing and transforms these data sources to comply with the D4UM data model. 
This data model formally specifies an integrated schema 
and establishes spatial, temporal and contextual connections across these sources,
thus facilitating integrated analytics going across the dataset boundaries.  
For example, traffic speed records are aligned with the street segments obtained from the OpenStreetMap data. 

The D4UM platform conducts long-term collection of mobility-related data from data streams to create an overview of relevant mobility data over longer time periods (i.e. months or years) within the \emph{long-term data collection} component.
In particular, recording of dedicated streaming sources 
such as traffic warnings published as RSS feeds or data extracted from social media channels (e.g. police channels on Twitter) allows for long-term analytics to observe patterns and temporal fluctuations, which would otherwise not be possible.
\emph{Data enrichment} facilitates collection of supplemental data by employing Web mining (e.g. by identifying events in social media streams) and citizen science tools (e.g. by collecting data with the MiC App presented in Section \ref{sec:mic}).

The \emph{data analytics} layer introduces models that build on the integrated data. 
For example, analysis of long-term data collections can be employed to identify typical urban movement patterns. For instance, spatio-temporal dependencies between traffic conditions on different roads can reveal structural problems within the road network \cite{isprs-archives-XLII-4-185-2018, DBLP:conf/gis/LiangJZ17}. Another example is the analysis of the impact of external factors (e.g. heavy rain or snowfall) on mobility behaviour \cite{7727607}. 
Data-based forecasts can make use of long-term data collections to predict urban mobility patterns in the future. For example, in the presence of planned special events (e.g. football matches or concerts), an increased load on the mobility infrastructure such as roads and public transportation capacity can be estimated \cite{DBLP:conf/gis/ZhouKLSZ16, 7765036}. 

Dedicated \emph{interfaces} such as APIs and graphical user interfaces grant access to the data analytics results. 
APIs serve as a conceptual abstraction from the complex models and provide information in a machine readable form, such as the GeoJSON\footnote{\url{https://tools.ietf.org/html/rfc7946}} format. 
Graphical interfaces such as map layers graphically present the analysis results and 
allow for embedding a map into Web pages. 

Finally the \emph{services, apps and pilots} layer makes the analysis results available for end users. In particular, we develop a dashboard Web application that can be used by city planers to analyse traffic patterns on the long term, while citizens can use the MiC app to derive insights about their own mobility behaviour and voluntarily contribute data to the D4UM system.

\section{Integration of Web-based Mobility Data}
\label{sec:data}

An overview of existing data sources from which we currently extract regional and other relevant information in the context of urban mobility is presented in Table \ref{tab:dataSources}. The majority of these data sources (except the traffic flow information) are Web-based. 
Traffic Data information sources include traffic flow information and traffic feeds. Traffic flow data, provided as aggregated Floating Car Data (FCD), reflects the average speed of the road traffic with respect to the individual road segments, whereas traffic feeds contain traffic warnings and accident notifications provided as RSS feeds.
Public Transport Information includes public transportation query logs and GTFS data.
We obtain query log data from the \emph{EFA}-system\footnote{\url{https://www.efa.de}}. \emph{EFA} is the official Web service for routing and time table information of public transportation for the region of Hannover. In this context, a query is a request for a public transportation route with specified origin, destination and departure time, issued via a Web-interface or a mobile application. 
Moreover, we consider Global Transit Feed Specification (GTFS)\footnote{\url{https://developers.google.com/transit/gtfs/}} data which provides timetable information for public transportation such as departure times, routes, etc.
This data is complemented with regional weather conditions, which can potentially impact the selection of the transportation mode and routes. 

Next to such directly related mobility data, we consider additional information obtained from social media and the Web. 
We consider social media data obtained from the Twitter streaming API\footnote{\url{https://developer.twitter.com/en/docs/tweets/filter-realtime/overview}}, which potentially contains information about regional events as well as traffic incidents.
Web data includes 
event-centric Web markup, which is prevalent in Web pages through standards such as RDFa\footnote{RDFa W3C recommendation: http://www.w3.org/TR/xhtml-rdfa-primer/}, and Microdata\footnote{http://www.w3.org/TR/microdata}. We currently investigate in particular data complying with schema.org as the most established markup vocabulary on the Web so far \cite{Meusel:2014}. 
In our previous work we developed methods to infer missing categorical information in noisy and sparse Web markup data \cite{tempelmeier2018}, increasing usefulness of this data for event-centric applications. 
Furthermore, we consider event-centric focused crawls from the Web \cite{Gossen:2015} and Web archives \cite{GossenDR17}, \cite{Gossen2018}, 
as well as Twitter data regarding events and traffic.
Another source of event-centric information is the recently proposed EventKG knowledge graph \cite{GottschalkD18,GottschalkSWJ}.
Finally, this information is complemented with geographic data (e.g. street networks, locations of event venues) obtained from \emph{OpenStreetMap}\footnote{\url{http://www.openstreetmap.org/}}.
The information contained in these sources is highly complementary. 

These data sources can be used to address parts of the afore introduced research questions. For instance, in the context of events, the traffic flow and public transportation query data can be used to determine typical patterns while Web, Web archives, knowledge graphs and social media sources can provide information about planned special events such as concerts or football matches. Another example is the computation of the average load on the roads with respect to the external factors, e.g. weather. 
These approaches require integration of data originating at several data sources, where semantic data descriptions, methods of dataset profiling \cite{EllefiBBDDST18}, \cite{DietzeDT19} and data quality analytics play an important role.

\begin{table*}[htb]
\fontsize{8pt}{9pt}\selectfont
\caption{An overview of regional and general mobility-relevant data sources for Lower Saxony, Germany.}
\label{tab:dataSources}
\centering

\begin{tabular}{|p{0.2\textwidth}|p{0.2\textwidth}|p{0.18\textwidth}|p{0.13\textwidth}|p{0.13\textwidth}|p{0.05\textwidth}|}
\hline
\textbf{Source}   & \textbf{Description}                                                            & \textbf{Granularity / Size}    &   \textbf{Timespan}                                 & \textbf{License}        & \textbf{Format} 
\\ 
\hline

\multicolumn{6}{l}{\textbf{Traffic Data}} \\
\hline
Traffic flow     & Average car traffic speed records in Lower Saxony, Germany.                                           & Average traffic speed per road segment and time interval (15 min) &
September 2017 - January 2019
& Commercial               & CSV             \\ \hline

Traffic feeds     & Traffic warnings and incidents\footnote{E.g. in Lower Saxony, Germany: \url{http://www.vmz-niedersachsen.de/}
}.                                                      & $\sim$ 25000 notifications                   & June 2017-January 2019
	& Provider-specific                 & RSS feeds  / XML         \\ 
\hline

\multicolumn{6}{l}{\textbf{Public Transport Information}} \\
\hline
Public transportation query logs        &  Query logs for public transportation routes and timetables.                                             & $\sim$ $4 \cdot 10^6$ queries per month  &
	October 2016-January 2019
    & Commercial              & CSV
\\ 
\hline
Global Transit Feed Specification (GTFS) data &  Public transportation timetable information for Lower Saxony, Germany. & 8800 stops, 2600 routes,  $2 \cdot 10^6$ stop times & until January 2019  & Open Data & CSV\\
\hline

\multicolumn{6}{l}{\textbf{City Data}} \\
\hline
Rainfall data & Volume of rain in a region. & 1 Record / hour / km$^2$
& 2005-2019
& Open data & CSV  \\ 
\hline

\multicolumn{6}{l}{\textbf{Social Media}} \\
\hline
Twitter           & Event- and location-centric tweets from Twitter API. Traffic information channels\footnote{E.g. from Hannover police and {\"U}stra (a public transportation provider) \url{https://twitter.com/Polizei\_H},  \url{https://twitter.com/uestra}}.               & 
German tweets from Twitter API &
May 2017-January 2019 & Twitter API license             & JSON            \\ 
\hline

\multicolumn{6}{l}{\textbf{Web}} \\
\hline
Event-centric Web markup            & Annotated Web pages, e.g. using schema.org. & Web Data Commons event subset: $263 \times 10^6$ facts  & until November 2017
& Common Crawl ToU
 & RDFa, MicroData \\ \hline
Focused crawls & Event-centric crawls, news\footnote{Collected using iCrawl toolbox \url{http://icrawl.l3s.uni-hannover.de/}}.                                    & $\sim$ 22000 events located in Hannover, Germany    & Crawl-specific
& Provider-specific       & HTML      \\ 

\hline

EventKG & Multilingual event-centric temporal knowledge graph & $690 \cdot 10^3$ events and $2.3 \cdot 10^6$ temporal relations (V1.1) &  until today & Creative Commons Attribution Share Alike 4.0 & RDF\\

\hline
OpenStreetMap     &  Geometries (points, lines, polygons) annotated with properties. Subset for parts of Lower Saxony.                                 & $2.6 \cdot 10^6$ facts & until January 2019
& ODbL               & XML             \\ 

\hline
\end{tabular}
\end{table*}

However, for comprehensive analysis of urban mobility some data is still missing. In particular foot walks, bicycle rides and intermodal data is not captured by any of the existing data sources. 
On the contrary, these modes become increasingly popular and thus important for urban mobility applications. To this end, the project introduces the MiC App, which can be used to collect the required data.

\section{MiC app}
\label{sec:mic}

To address the lack of data regarding foot walks, bicycle rides and intermodal trips, the project introduces the \emph{Move in the City} (MiC) app. This app poses a citizen science inspired method to gather movement data. Users of the MiC app can capture GPS traces of their trips with their mobile devices and voluntarily contribute the data to the D4UM project. In return, individual movement statistics are provided to the user. Moreover, the increased amount of available mobility data enables novel analytics of mobility behaviour. Based on the analytics, previously unseen problems can be identified and addressed, creating further benefits for the user and for the mobility ecosystem as a whole. 

Movement data captured by the MiC app undergoes a preprocessing routine tailored to the specific needs of urban mobility analytics. Different modes of mobility such as walking, cycling, driving and the use of public transportation are automatically distinguished. Furthermore, if public transportation is used, the data is enriched with additional information such as entry stop, exit stop and the public transportation line used. This way, the data captured by the MiC app represents a valuable data source for further mobility analytics approaches. 

\tikzset{component/.style={draw, align=center, minimum width=35mm, minimum height=15mm,
		 rounded corners=1mm}}
\tikzset{point/.style={circle,inner sep=0pt,minimum size=0pt}}
\tikzset{database/.style={shape border rotate=90,draw,
		 minimum size=18mm}}
\begin{figure}
\begin{center}
\scalebox{0.65}{
\begin{tikzpicture}[node distance=5mm and 12mm]
\node (app) [component] {Mobile Client};
\node (MQTT) [point, right=of app] {};
\node (Client) [component, right=of MQTT] {Database\\Input Service};
\node (Data) [component, below=of Client] {Data Processing\\Service};
\node (API) [component, below=of Data] {API Service};
\node (DB) [database, right=of Data] {Database};
\node (HTTP) [point, left=of API] {};
\node (HTTP2) [point, left=of Data] {};
\node (Viz) [component, left=of HTTP] {Analytics\\Applications};
\node (EFA) [component, left=of HTTP2] {Public Transportation\\Information System};
\draw[->, very thick] (app) to node[above] {} (Client);
\draw[->, very thick] (Client) to[bend left=32] (DB);
\draw[<->, very thick] (Data) to (DB);
\draw[->, very thick] (DB) to[bend left=32] (API);
\draw[->, very thick] (API) to node[above] {} (Viz);
\draw[->, very thick] (EFA) to node[above] {} (Data);
\end{tikzpicture}}
\caption[Component overview]{Overview of the MiC app data flow.}
\label{fig:components}
\end{center}
\end{figure}
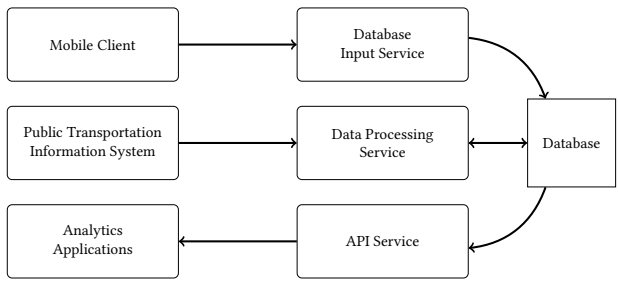

\subsection{Architecture}
The architecture of the MiC app aims at keeping a low computation effort on the user device to reduce the energy consumption of the app. Furthermore, the architecture is platform independent to enable as many users as possible to use the app. To this end, MiC makes use of established, flexible Web technologies such as Web-based user interfaces and the MQTT protocol\footnote{http://mqtt.org/}. To further ensure platform independence, the app does not directly access sensor data of the mobile devices. Instead, the relatively new, built-in activity recognition APIs of the mobile operating systems are used. In addition to the activity recognition data, the app records fine grained location data.

\begin{figure*}
    \begin{subfigure}[c]{0.27\textwidth}
        \includegraphics[width=\textwidth]{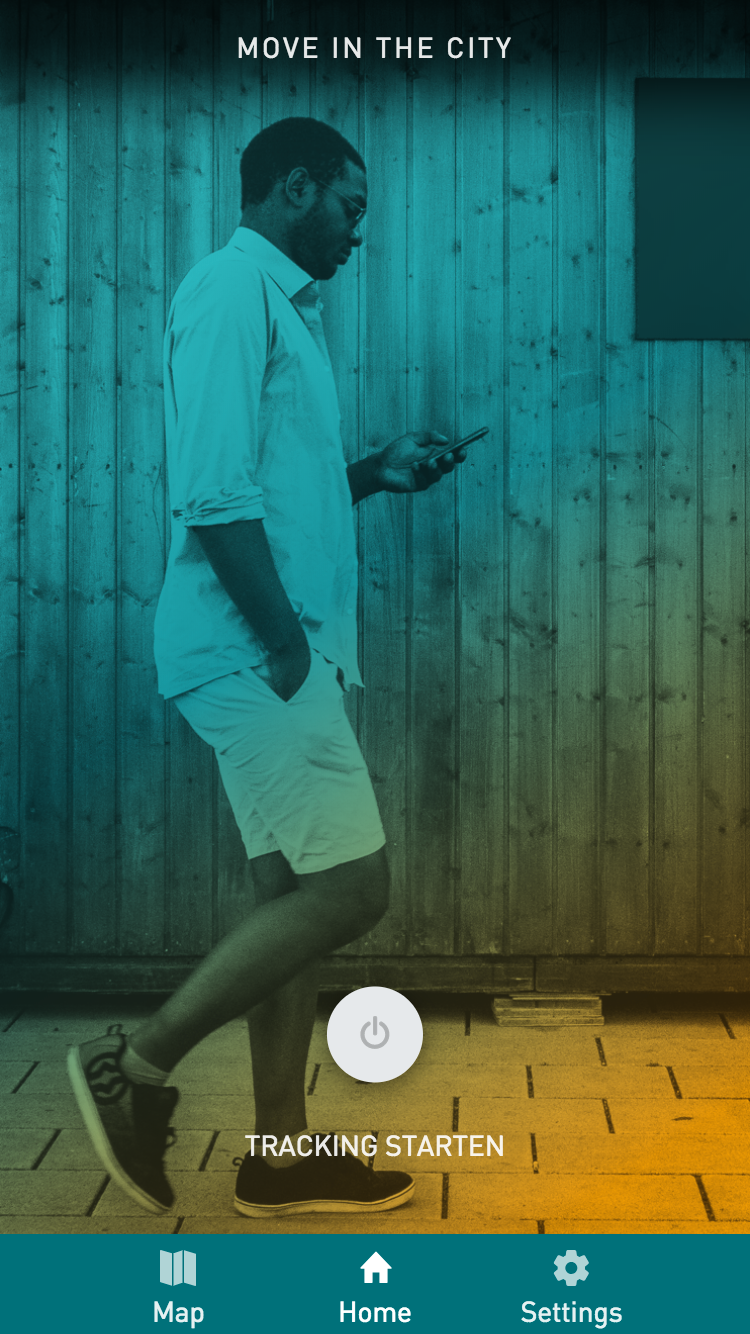}
        \caption{Start View}
        \label{fig:mic_screens_start}
    \end{subfigure}
    \begin{subfigure}[c]{0.27\textwidth}
        \includegraphics[width=\textwidth]{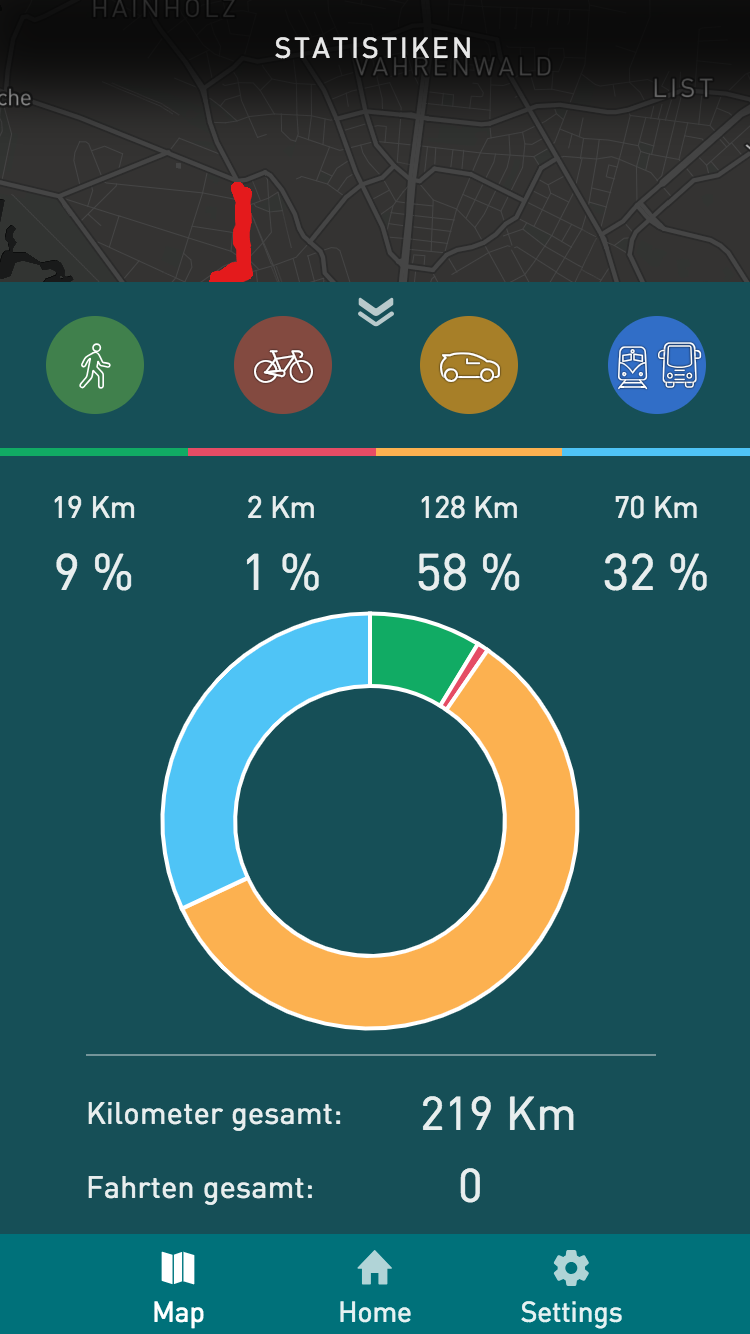}
        \caption{User Statistics View}
           \label{fig:mic_screens_stats}
    \end{subfigure}
    \begin{subfigure}[c]{0.27\textwidth}
        \includegraphics[width=\textwidth]{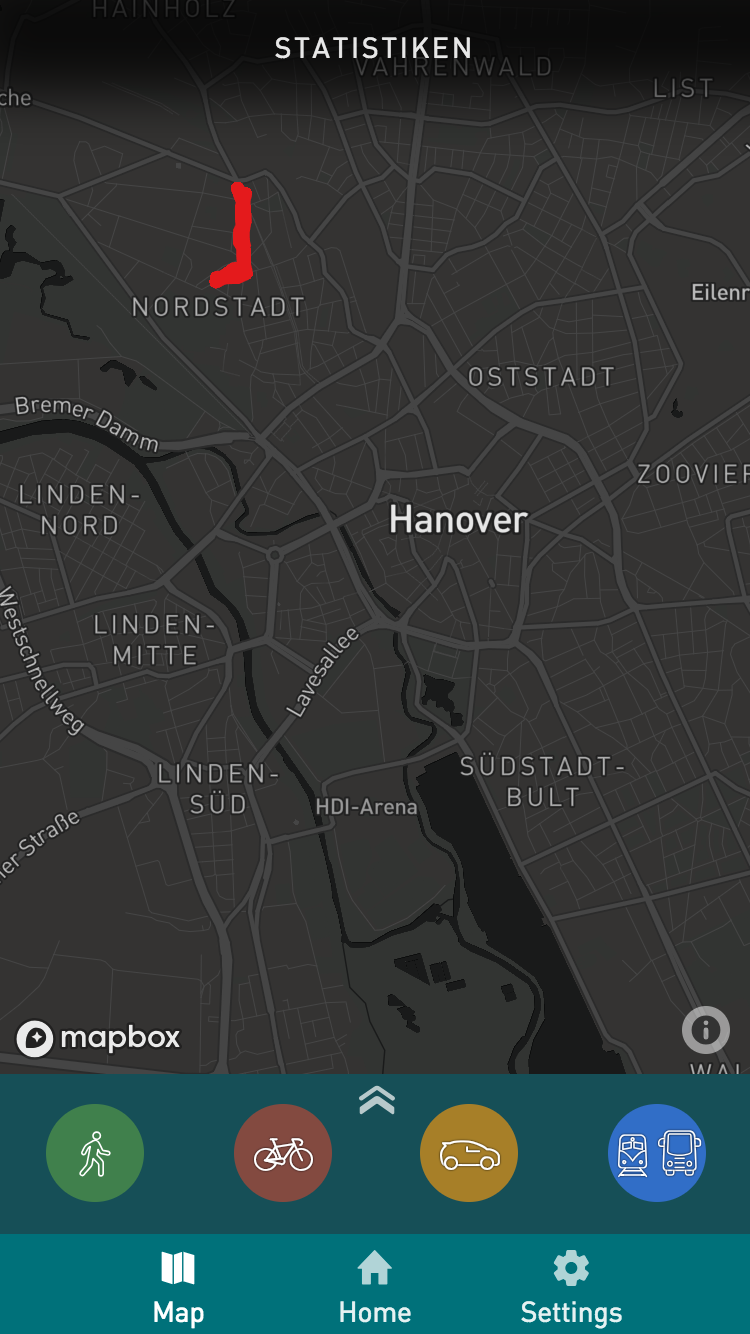}
        \caption{Map View}
        \label{fig:mic_screens_map}
    \end{subfigure}
    \caption{Screenshots taken from the MiC app. Subfigure (a) presents the start view. Users can start recording their movement by pressing a single button. Subfigure (b) presents the user statistics view. It visualises details about the user specific mobility behaviour, e.g. the user specific fraction of transportation modes. Subfigure (c) presents the map view that projects captured trip data onto a map.}
        \label{fig:mic_screens}
\end{figure*}

Figure \ref{fig:components} provides an overview on the data flow. Data recorded by the mobile client is sent to a Web-based endpoint which stores the recorded data in a database. The data is then enriched with information about public transportation, obtained by querying the local information system for public transportation. Finally, a Web-based endpoint provides an API that makes the data accessible for data analytics applications.

\subsection{App Design}

MiC aims at motivating the users to use the app by providing an appealing and easy to use interface. Figure \ref{fig:mic_screens} presents two exemplary screenshots of the MiC app. The start view can be seen in Figure \ref{fig:mic_screens_start}. We achieve a lightweight user interaction by only requiring users to press one button to start or stop the recording of their movement. Figure \ref{fig:mic_screens_stats} depicts the view presenting statistics of the individual user. The view presents the fraction of the user transportation modes by providing absolute and relative numbers as well as a graphical visualisation. Ideas for further user statistics include visualisation of the environmental impact of the users mobility behavior or the overall contribution to the urban mobility network. Figure \ref{fig:mic_screens_map} presents the map view, where the user can visualise captured trip data, i.e. GPS traces, on a map.

\subsection{Piloting}

The first closed test of the system was conducted by five participants who made 41 trips over the duration of five days, where the median duration of the recorded trips took 24 minutes.
The participants wrote detailed travel diaries for these trips that were then used as a ground truth to assess the accuracy of the mode of travel classification. Due to limited GPS signals paths in underground trains, they were excluded from the study.

Table \ref{tab:accuracy} presents the median duration and accuracy of transportation mode recognition of the recorded trips with respect to mode of transportation, where the total row summarises all recorded trips. Trips that include multiple modes of transportation were divided into individual trips with only one mode of transportation. In addition to classifying the correct  transportation mode, we only considered tram and bus trips to be recognised correctly, if the correct entry stop, exit stop and public transportation line was identified by the system.

In total, 86.2\% percent of the trips were recognized correctly by the system.
The highest accuracy was achieved for the recognition of bicycle trips (100\%). This is due to the relatively clear movement signature of riding a bicycle, which is well recognised by the activity recognition of the smartphones.
The second highest accuracy is achieved for cars.
Even though trams and busses achieve the lowest accuracy, the absolute accuracy is at least 73\%. The reduced accuracy for the two classes is likely to be caused by the additional constraints, i.e. the recognition of start stop, entry stop and public transportation line.

\begin{table}
\centering
\caption{Median duration and accuracy of mode of travel recognition with respect to transportation mode. Trips including multiple modes of transportation are divided into individual trips with only one mode of transportation.}
\begin{tabular}{ccccc}
\toprule
\textbf{Mode} & \textbf{Number} & \textbf{Median Duration} & \textbf{Accuracy} \\
\hline
Bicycle & 16 & 10 min. & 100\% \\
Car & 14 & 9.5 min. & 92.9\% \\
Tram & 13 & 7 min. & 76.9\% \\
Bus & 15 & 12 min. & 73.3\% \\
Total & 58 & 11 min. & 86.2\% \\
\bottomrule
\end{tabular}

\label{tab:accuracy}
\end{table}

The first public test phase of the system is currently ongoing. The application beta testing platforms by Apple and Google are being used to carry out the application. Users were recruited at universities and expositions. Table \ref{tab:micstas} presents statistics about data currently captured by the MiC app. To today, 78 participants were found to test the system under real-world conditions.

\begin{table}[]
    \centering
    \caption{Statistics of data currently captured by the MiC app.}
    \begin{tabular}{lr}
        \toprule
        \textbf{Property} & \textbf{Value} \\
        \midrule
         Number of Users & 78  \\
         Number of Recorded Trips & 218 \\
         Average Trip Duration & 69 min. \\
         Number of Captured GPS Points & 92.550 \\
         \bottomrule
    \end{tabular}
    \label{tab:micstas}
\end{table}

\subsection{Data Protection}
To ensure data protection compliant to latest regulations (e.g. the \emph{General Data Protection Regulation} by the European Union), a concept was developed in joint work with researchers from legal informatics that includes the following components: 

\begin{itemize}
\item The consent of app users (Article 6 GDPR); 
\item A privacy policy (Article 13 GDPR); 
\item Data pseudonymisation (Article 89 GDPR);
\item The agreement between joint controllers (Article 26 GDPR).  
\end{itemize}

The MiC app has an integrated privacy policy and consent form providing app users with the information about the data procession, data rights and possibility to consent. It 
contains details about the types of data the MiC app collects, the collection procedure, purposes of research, data storage and data sharing within the D4UM project, data
rights of the users (i.a. access, deletion, withdrawal of consent) and contact details of the app providers.

The geo-referenced data collected by the MIC app is processed in pseudonymised form. By this, the data is purified from all personal identifiers, such as name, e-mail, IMEI number (a cellphone serial number), etc. 
Pseudonymisation is chosen as a de-identification measure (in contrast to anonymisation or irreversible de-identification) to enable enforcement of the data subjects' rights to 
access and deletion of the data, as may be requested by the users. This is part of an integrated and innovative citizen science approach which enables everyone to actively participate in research and planning processes with a strong effort to establish new standards for data sovereignty of citizens and cities.

\section{Use Cases}
\label{sec:use_case}
In this section we discuss two exemplary use cases for the D4UM platform and the MiC app.

\subsection{Urban Mobility in Presence of Planned Special Events}
Large-scale planned special events such as football games, concerts, etc. are known to have impact on both road traffic \cite{KWOCZEK2014973} and public transportation services \cite{7765036}. 
Analysing mobility behaviour in such situations is particularly challenging due to a great number of influence factors (e.g. target audience, weather conditions, public transportation infrastructure). Moreover, large-scale events typically have an impact on all modes of transportation, which are mutually dependent. Thus, a holistic approach, which considers data about all modes of transportation as well as external factors, is required to gain a better understanding of the mobility behaviour in presence of planned special events.

\begin{figure*}
    \centering
    
    \begin{subfigure}{0.40\textwidth}
    \centering
        \includegraphics[width=\textwidth]{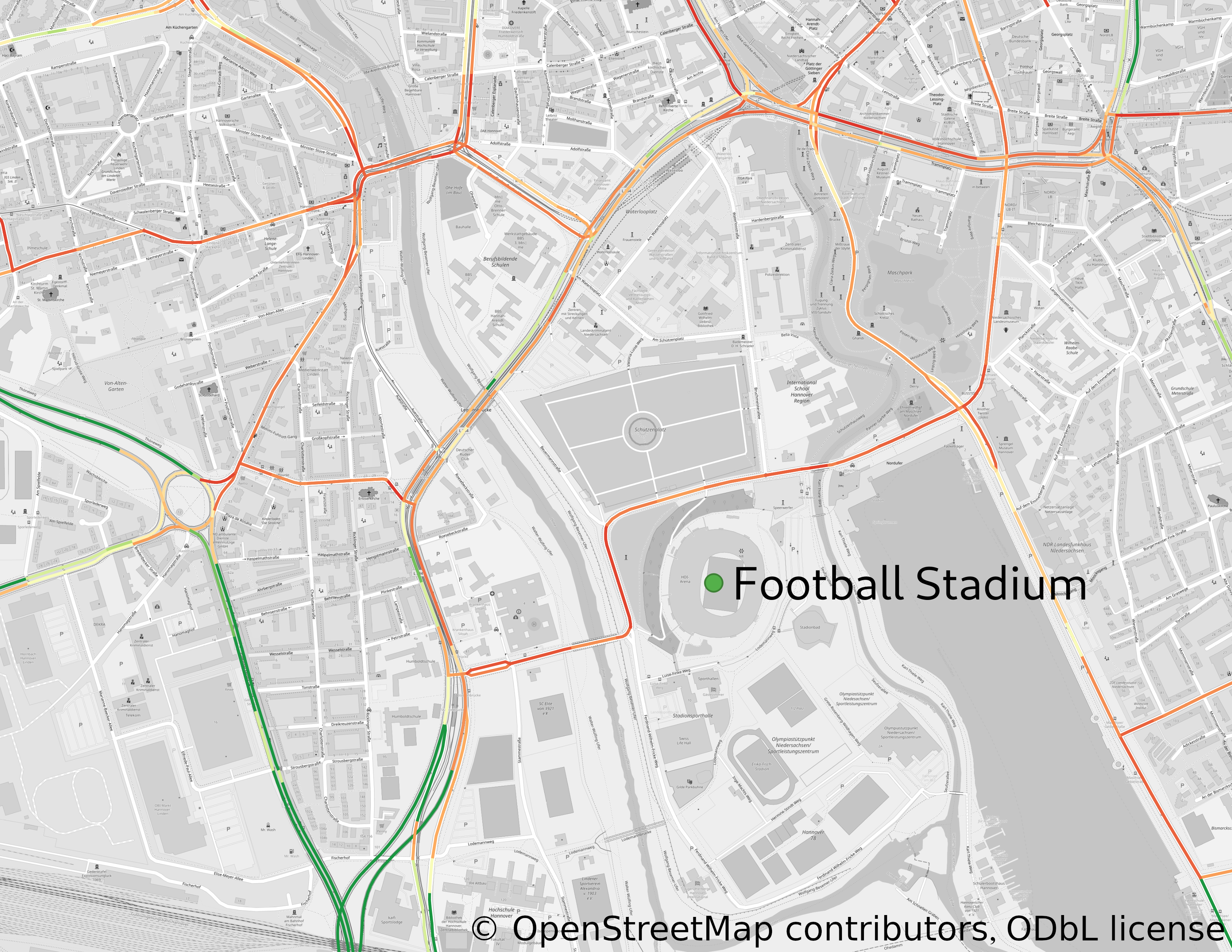}
        \caption{Traffic conditions on the major roads located nearby the football stadium half an hour before the game starts. Red color indicates heavy load, green color indicates low load.}
        \label{fig:footballRoads}
    \end{subfigure}
    \hfill
    \begin{subfigure}{0.48\textwidth}
        \includegraphics[width=\textwidth]{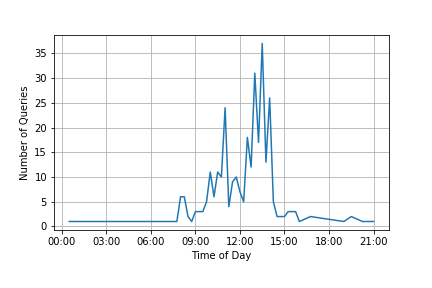}
        \caption{Number of public transportation queries for the bus stop nearest to the football stadium with respect to the time of day on the day of the football game.}
        \label{fig:footballQueries}
    \end{subfigure}
    \caption{Illustration of an impact of a football game on road traffic and public transportation. The football game took place on December 17th, 2017 at 3:30pm in Hannover, Germany.}
    \label{fig:football}
\end{figure*}

We illustrate the complexity of the problem at the example of a football game that took place in the city of Hannover (Germany) on December 17th, 2017 at 3:30pm.
Figure \ref{fig:football} presents the conditions around the football stadium in Hannover for both road traffic and public transportation services. Figure \ref{fig:footballRoads} depicts the road traffic conditions around the stadium half an hour before the start of the game. 
As we can observe, the roads with a high load (red color) as well as the roads with good traffic conditions (green color) are present nearby the stadium. This illustrates that the impact of planned special events might be complex and such impact does not necessarily evenly spread around the event venue, but might be subject to other factors such as road topology or availability of parking spaces. 
Figure \ref{fig:footballQueries} presents the number of queries to the public transportation information system that were issued for the bus stop near the stadium during the day of the football game. 
We observe a relatively low number of queries during the morning hours. 
At 9am, the number of queries starts to rise and reaches their maximum at 2:30pm, one hour before the game starts. We assume that an increased number of queries in the temporal proximity of the event is likely related to the football game. 
Overall, we observe, that a single factor, i.e. the taking place of a large-scale event, can impact both road traffic and public transportation simultaneously. Therefore, analysis of these effects should take both mobility modes into account.
To this extent, the D4UM platform presented in this paper 
aims at holistic data analysis of urban mobility data.

\subsection{Analytics of Urban Mobility Infrastructure}
%

The rapid growth of cities has lead to an increased demand of suitable urban mobility infrastructure. New services such as car sharing, (e-)bike sharing and ride sharing have emerged and begin to gain importance within the mobility services ecosystem. However, data about these modes of transportation is typically only sporadically available. Moreover, intermodal data which captures different modes of transportation on a single journey is evenly rare. 

%
%
%

The interdependencies between individual and public transportation increase the complexity of the problem. For instance, a poor coverage of public transportation services might lead to an increased load on the road infrastructure. In turn, an overloaded road system can lead to an increased demand of public transportation services or bicycle tracks.

Therefore it is crucial for city planers and mobility service providers to conduct holistic analysis, which take all modes of transportation into account. Such analysis should include the identification of:
\begin{itemize}
    \item typical mobility patterns
    \item demand of mobility services
    \item interdependencies between transportation modes
    \item coverage of public transportation services
\end{itemize}

The D4UM platform enables the holistic analysis of urban mobility data in these scenarios. Furthermore, the MiC app presented in this paper is a valuable approach to capture rarely available data about urban mobility behaviour, including bicycle rides and intermodal data and complement available data sources.

\section{Related Work}
\label{sec:background}
In this section we discuss related work in the area of smart city mobility systems as well as approaches to predict individual aspects of urban mobility.






\subsection{Smart City Mobility Systems}
\cite{Moustaka:2018:SRS:3271482.3239566} conducted a systematic review of smart city data analysis approaches and provide taxonomies for data sources, data analytic methods and smart city services.
Urban CPS \cite{Zhang:2015:UCS:2735960.2735985} is a cyber physical system that integrates floating car data, cellphone data and public transportation data to make prediction about real time traffic speeds.
\cite{Lecue:2014:SST:2557500.2557537} employ semantic technologies to develop STAR-CITY, a system for 
traffic prediction and reasoning, used for spatio-temporal analysis of the traffic status as well as for the exploration of contextual information such as nearby events.
\cite{10.1007/978-3-642-40994-3_50} presents a system that makes use of cellphone data to analyse the demand of public transportation services within a city.
While these approaches focus on predictions for one mode of mobility only and mainly use a single dataset as their primarily data source, we consider the other modes of transportation as well, e.g. bicycle rides or public transportation.
\cite{5673950} provides a summary of common enterprise, logical and physical architectures for digital smart city applications. While the authors consider architectures for general smart city applications, we focus on the mobility domain.
\cite{4511428} proposed the Compressed Start-End Tree (CSE-tree), a spatio-temporal index structure that can be used to effectively index and retrieve temporal GPS data which is in particular relevant for smart city mobility applications. 

\subsection{Urban Mobility Analytics}
Recently a number of studies addressed several individual prediction tasks at the interface of the urban infrastructure and mobility. We consider these approaches as potential use cases for the D4UM platform.

\textbf{Traffic Forecasting:}
Short-term forecasting of urban road traffic has been the focus of numerous studies where data sparsity is a particular challenge.
\cite{Wang:2014:TTE:2623330.2623656} tackles the problem by using sparse FCD (floating car data) and a context-aware tensor decomposition approach to estimate travel times for road segments for which no FCD is available. The information is then used to estimate the required travel time for a given route. 
\cite{DBLP:conf/gis/MengYSG017} proposed a framework for the city-wide inference of traffic volume. They make use of a semi-supervised learning algorithm that can be used with sparse loop detector data as well as taxi GPS data.
Similar, \cite{10.1371/journal.pone.0145348} employs a hidden Markov model to estimate traffic speeds of a road network based on sparse FCD where the speed to be estimated on a single road is considered as a hidden state.

\textbf{Social Network Data:}
Further approaches utilized location-based social network (LBSN) data.
\cite{doi:10.1111/tgis.12289} leverages LBSN data to identify functional urban regions. They employ latent Dirichlet allocation and unsupervised machine learning algorithms to determine the regions.
\cite{Li2018} investigated the general predictability of LBSN data. They conducted a case study on Foursquare datasets, where users indicate their geographic location, i.e. the users can indicate that they are at a certain event venue. The authors do not focus on a specialised prediction task, but provide general insights on working with the aforementioned data.
\cite{doi:10.1080/13658816.2017.1282615} make use of LBSN data to infer boundaries of functional regions in urban environments. They construct a mobility network from spatial user interactions and delineate boundaries by identifying strongly connected communities within the network space.
\cite{7583675} detects events from social media by employing a hashtag-based algorithm. 
The event information extracted from the social media is then used 
for prediction of the public transportation flow.

\textbf{Structural Analysis:}
Another class of approaches focuses on discovering structural dependencies within cities by analysing urban mobility data.
\cite{Anwar:2016:TEC:2983323.2983688} propose a
method to keep track of the congestions in urban road networks to identify unstable road segments. \cite{Jin:2016:CCF:2983323.2983655} make use of context-aware tensor
decomposition to identify so called \emph{urban black holes}, 
i.e. traffic anomalies with a greater inflow than outflow.
\cite{Hong:2015:DUB:2820783.2820811} detect urban black holes by using a grid-based index-structure that is build on top of a spatio-temporal graph representing the road network. They extract candidate cells from the index which then are used to determine the exact subgraphs that are urban black holes.
\cite{DBLP:conf/gis/LiangJZ17} identify cascading patterns on congested roads. They propose a generative probabilistic model that maximises the likelihood of a cascade to be present with respect to the observed traffic data.
\cite{DBLP:journals/gis/KempinskaLS18} employs topic modelling to analyse urban street works. They proposed the concept of interactional regions, i.e. regions that commonly bound routes within the street networks. 
\cite{Wang:2017:RRL:3132847.3133006} leverage taxi flow data to learn vector representation of city regions. The representations are then used to make predictions about the regions such, e.g. crime rate, average income or average house prices.
\cite{10.1371/journal.pone.0119044} employ deep learning techniques, i.e. a combination of restricted Boltzmann machines and recurrent neural networks, to learn high-dimensional congestion patterns from taxi GPS data. 
\cite{6266748} make use of taxi GPS-trajectories to classify the land use of urban areas. They propose an iterative DBSCAN algorithm to cluster regions with respect to the frequency with which passengers are picked up or set down. They make use of the same information to classify the land use of regions, e.g. the land use for hospitals or commercial districts.
\cite{Zhang:2018:SUR:3219819.3219987} proposes the infinite urbanization process model that employs a topic modelling approach to simultaneously discover the function of an urban region (i.e. the distribution of present shops, restaurants, etc.) and to estimate the region popularity (e.g. in terms of real estate prices).
Similar, \cite{Sun:2018:EUR:3219819.3220009} extracts urban regions of interest from online map search queries. They propose a spatio-temporal latent factor model which identifies travel patterns that influence points of interest.

\textbf{Special Traffic Conditions}
Several approaches target the identification of problematic segments and areas of urban networks under specific conditions (e.g. planed special events).
%
\cite{DBLP:conf/itsc/KwoczekMN15} propose the use of an artificial neuronal network to identify road segments that are typically affected by planned special events that take place at a particular venue. 
\cite{DBLP:conf/gis/ZhouKLSZ16} proposed an approach to detect events from traffic data. They make use of a two-dimensional grid to partition the space. The authors then infer a graph-based representation of the grid that captures the flow of vehicles between the individual cells of the grid where the root of the graph is located at the events location.
%
Finally, \cite{7765036} investigate the effect of public events on the public transportation network. They propose a Bayesian additive model that can be employed to gain an understanding of public transportation demand in the presence of events. I.e. the model is able to predict the number of public transportation trips to the venues where the respective event takes place.


\section{Conclusions \& Future Work}
\label{sec:conclusion}

In this paper we presented our ongoing work towards holistic urban data analytics conducted in the context of the Data4UrbanMobility project. 
We presented the D4UM platform 
that facilitates seamless long-term analytics of heterogeneous mobility-related data sources, including but not limited to floating car data, weather conditions, traffic warnings and Web queries. Furthermore, we presented the MiC app - a citizen science application that facilitates complementing this data with 
intermodal mobility patterns of city inhabitants. 
In our future work we intend to further increase the MiC app 
user base, and implement further use cases on top of the integrated D4UM platform.

\section{Acknowledgements}
This work was partially funded by the Federal Ministry of Education and Research (BMBF), Germany, "Data4UrbanMobility" project, grant ID 02K15A040.

\balance

\bibliographystyle{ACM-Reference-Format}
\bibliography{ref}

\end{document}